\documentclass[aps,prb,superscriptaddress,twocolumn]{revtex4}

\usepackage{bm}
\usepackage{graphicx}
\usepackage{amsmath}

\begin{document}

\title{Extracting signatures of quantum criticality\\
       in the finite-temperature behavior of many-body systems}
\author{Alessandro Cuccoli}
%\email{cuccoli@fi.infn.it}
\affiliation{Dipartimento di Fisica, Universit\`a di Firenze,
    e Unit\`a CNISM,
    Via G. Sansone 1, I-50019 Sesto Fiorentino (FI), Italy}
\author{Alessio Taiti}
%\email{taiti@fi.infn.it}
\affiliation{Dipartimento di Fisica, Universit\`a di Firenze,
    Via G. Sansone 1, I-50019 Sesto Fiorentino (FI), Italy}
\author{Ruggero Vaia}
%\email{ruggero.vaia@isc.cnr.it}
\affiliation{Istituto dei Sistemi Complessi -- CNR,
via Madonna del Piano 10, I-50019 Sesto Fiorentino (FI), Italy}
\author{Paola Verrucchi}
%\email{paola.verrucchi@isc.cnr.it}
\affiliation{Centro di Ricerca e Sviluppo SMC
 dell'Istituto Nazionale di Fisica della Materia -- CNR,
 Sezione di Firenze, via G.~Sansone 1,
 I-50019 Sesto Fiorentino, Italy}
\affiliation{Dipartimento di Fisica, Universit\`a di Firenze,
    Via G. Sansone 1, I-50019 Sesto Fiorentino (FI), Italy}
\date{\today}

\begin{abstract}
We face the problem of detecting and featuring footprints of quantum
criticality in the finite-temperature behavior of quantum many-body
systems. Our strategy is that of comparing the phase diagram of a
system displaying a $T\,{=}\,0$ quantum phase transition with that of
its classical limit, in order to single out the genuinely quantum
effects. To this aim, we consider the one-dimensional Ising model in
a transverse field: while the quantum $S\,{=}\,1/2$ Ising chain is
exactly solvable and extensively studied, results for the classical
limit ($S\,{\to}\,\infty$) of such model are lacking, and we supply
them here. They are obtained numerically, via the Transfer-matrix
method, and their asymptotic low-temperature behavior is also derived
analytically by self-consistent spin-wave theory. We draw the
classical phase-diagram according to the same procedure followed in
the quantum analysis, and the two phase diagrams are found
unexpectedly similar: Three regimes are detected also in the
classical case, each characterized by a functional dependence of the
correlation length on temperature and field analogous to that of the
quantum model. What discriminates the classical from the quantum case
are the different values of the exponents entering such dependencies,
a consequence of the different nature of zero-temperature quantum
fluctuations with respect to thermal ones.

\end{abstract}

% 05.70.Jk - Critical point phenomena
% 73.43.Nq - Quantum phase transitions
% 75.10.Hk - Classical spin models
% 75.10.Jm - Quantized spin models
% 75.10.Pq - Spin chain models
% 75.40.-s - Critical-point effects, specific heats, short-range order
% 75.40.Cx - Static properties

\maketitle

\section{Introduction}

One of the most fascinating aspects of many-body systems is the
possible occurrence of a phase transition, either at finite or at
zero temperature. This latter case is generally referred to as a
genuine \emph{quantum} phase transition (QPT), by this meaning that
it is exclusively observed in quantum many-body systems. Zero-point
quantum fluctuations are recognized as the fundamental ingredient of
a QPT, in the same sense as thermal fluctuations are in ordinary
finite-temperature phase transitions: Whenever fluctuations are
frozen, as in the $T=0$ classical case, no phase transition may
possibly occur.

In the literature, a QPT is also commonly said to occur when, for a
given value of one of the Hamiltonian parameters, the ground state of
the model qualitatively changes its structure. This definition is not
quite rigorous, firstly because it labels QPT any change in the
universality class of the model, as well as mean-field phenomena such
as saturation, secondly because it may paradoxically be extended to
classical systems: As a matter of fact, qualitative changes in the
structure of the minimum-energy configuration may well be observed,
for a given value of some Hamiltonian parameter, also in classical
systems at $T=0$, despite fluctuations being frozen. In this
framework, therefore, the comparison between the behavior of a
quantum system displaying a QPT and that of its classical limit
becomes meaningful even at zero temperature, as shown in the next
Section.

When temperature is switched on, the very same definition of QPT
looses its meaning; however, it is well established that a QPT
induces a peculiar finite-temperature behavior which is qualitatively
described by the best known phase diagram introduced in
Ref.~\onlinecite{ChakravartyHN1989}, and reported in
Fig.~\ref{f.phd}, where three different regimes appear
(\emph{renormalized classical}, \emph{quantum critical}, and
\emph{quantum disordered}), separated by crossover lines thoroughly
discussed in Ref.~\onlinecite{Continentino2001}. The relevance of
this phase diagram is mostly due to its suggesting that signatures of
a genuine quantum critical behavior may survive also at finite
temperature, a fact that opens the possibility to observe them
experimentally. Moreover, a renewed interest has arisen since
entanglement properties have entered the physics of many-body
systems, and questions like ``how resistant to thermal noise are
certain quantum properties?'' became essential in order to test
possible realizations of quantum devices.

Despite the above phase diagram being considered as strictly peculiar
to quantum systems, especially as far as the quantum critical and
quantum disordered regimes are concerned, its structure results from
the interplay between thermal (\emph{classical}) and \emph{quantum}
fluctuations: A precise analysis of the role played by these two
components is therefore necessary in order to ascertain whether the
latter play an essential role or not, and to distill genuine
footprints of quantum criticality to be experimentally looked for. To
this end, knowing the behavior of the quantum system at finite
temperature is not enough, and a careful comparison with the
corresponding classical limit is necessary. The quite unexpected lack
of results for the classical limit of quantum models displaying a QPT
has made such comparison unavailable until now.

This paper is aimed at filling this gap: We consider one of the
paradigmatic models displaying a QPT, namely the one-dimensional
quantum Ising model in a transverse field (QIF), and compare its
behavior with that of its classical limit, namely the classical Ising
model in a transverse field (CIF), which we study numerically, via
the Transfer-matrix method, and analytically, via self-consistent
spin-wave theory. The result is unexpected: A finite temperature
phase diagram is disclosed also for the CIF, and it has the same
structure of that for the QIF. In full analogy with the quantum case,
we identify three different regimes on the basis of the field and
temperature dependence of the correlation length; moreover, we show
that the algebraic behavior, which was thought to specifically
characterize the quantum critical and quantum disordered regimes,
does in fact show up also in the classical system, though with
different exponents. This result jeopardizes the experimental
renderings based on the statement that the observation of an
algebraic dependence of the correlation length implies the occurrence
of a quantum critical or disordered
regime~\cite{BitkoRA1996,RoennowMH1999,CCCMRTV2000,RoennowPJARM2005}.
In fact, an accurate analysis of the exponents is here shown to be
necessary in order to discriminate genuine quantum effects.

The structure of the paper is as follows: in Sec.~\ref{s.ItF.T=0} we
introduce the model and discuss its zero-temperature behavior. In
Sec.~\ref{s.ItF.T>0}, we first summarize the known results for the
QIF, and then present our results for its classical limit, the CIF:
numerical data for the magnetization and the susceptibility in the
field direction, and for the specific heat are shown and discussed.
The analysis of the field and temperature dependence of the
correlation length is considered in Sec.~\ref{s.PhaseDiagram}, where
the classical phase diagram is finally obtained and compared with
that for the QIF. Conclusions are drawn in Sec.~\ref{s.concl}.

\section{The Ising model in a transverse field: $T=0$}
 \label{s.ItF.T=0}
  \subsection{The quantum model}
   \label{ss.QIF.T=0}

\begin{figure}
\includegraphics[height=84mm,angle=90]{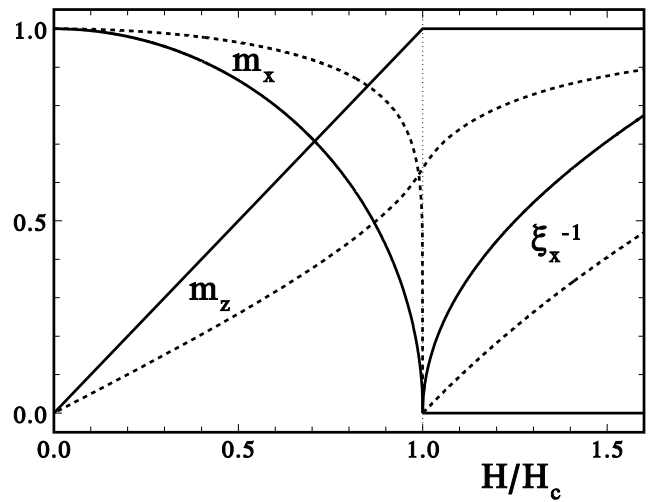}
\caption{
Zero temperature limit of $m_x$, $m_z$, and $\xi_x$ of the classical
(solid lines) and of the quantum (dashed lines) Ising model in a
transverse field.
}\label{f.zerot}
\end{figure}

One of the best known examples of a many-body system displaying a
QPT~\cite{Sachdev1999,Continentino2001} is the one-dimensional
quantum Ising model in a transverse field (QIF), whose Hamiltonian
reads
\begin{equation}
 \frac{\hat{\cal H}}{J}=-\sum_{i}
 \big(\hat{S}^x_i\hat{S}^x_{i+1}+H\,\hat{S}^z_i\big) ~,
 \label{e.HQIF}
\end{equation}
where $i$ runs over the sites of an infinite chain, and
$\hat{\bm{S}}_i$ are $S{=}1/2$ spin operators, $J$ is the exchange
energy constant and $H$ is the transverse field in units of $J$; this
model is exactly solvable~\cite{Katsura1962,Pfeuty1970} by means of a
Jordan-Wigner transformation to Fermi operators, and displays a QPT
at $T=0$ and $H=H_{\rm{c}}\,{=}\,1/2$: The discrete
$\hat{S}_i^x\to-\hat{S}_i^x$ global symmetry of the Hamiltonian is
spontaneously broken for $H<H_{\rm{c}}$, where the order parameter
$m_x=\langle\hat{S}_i^x\rangle/S$ becomes nonzero; long-range order
sets in at the critical point, as testified by the divergence of the
order-parameter correlation length. In particular, it is

\begin{equation}
  m_x\sim (H_{\rm{c}}-H)^{\beta}~~~{\rm for}~H\to H_{\rm{c}}^-~,
 \label{e.Mxquantum}
\end{equation}
and
\begin{equation}
  \xi_x\sim (H-H_{\rm{c}})^{-\nu}~~~{\rm for}~H\to H_{\rm{c}}^+~,
 \label{e.Xixquantum}
\end{equation}
with the exponents $\beta=1/8$ and $\nu=1$, as in the
finite-temperature phase transition of the classical two-dimensional
Ising model.~\cite{Onsager1944} Together with the other usual
exponents they obey typical scaling relations which, due to the
intrinsically dynamical nature of quantum fluctuations,
entail~\protect\cite{Continentino2001} the dynamical critical
exponent $z$. On the other hand, the magnetization along the field
direction, $m_z=\langle\hat{S}_i^z\rangle/S$, is an analytic function
of the field and changes its curvature at $H_{\rm{c}}$, where the
uniform susceptibility consequently displays a maximum. These
$T\,{=}\,0$ behaviors are reported as dashed lines in
Fig.~\ref{f.zerot}.

  \subsection{The classical model}
   \label{ss.CIF.T=0}

The classical limit of any given quantum system is unique, although
the converse is not true. For a spin system the limit $\hbar{\to}0$
must be taken while keeping finite the spin angular momenta
$\hbar\hat{\bm{S_i}}$.

From Eq.~\eqref{e.HQIF} one thus obtains the Hamiltonian of the
classical Ising model in a transverse field (CIF),
\begin{equation}
 \frac{\cal H}{J_{\rm{c}}}
 =-\sum_{i=1}^N\big(s^x_is^x_{i+1}+2h\,s^z_i\big) ~.
 \label{e.H}
\end{equation}
where $\bm{s}_i=(s_i^x,s_i^y,s_i^z)$ are \emph{classical spins},
i.e., three-dimensional vectors of fixed length $|\bm{s}_i|^2=1$,
while~\cite{OitmaaC1981}
\[
 J_{\rm{c}}=\mathop{~\lim~}
 \limits_{\stackrel{\scriptstyle\hbar{\to}0}{S{\to}\infty}} JS^2~,
 \hspace{10mm}
 2h =\mathop{~\lim~}
 \limits_{\stackrel{\scriptstyle\hbar{\to}0}{S{\to}\infty}} \frac HS
\]
are the classical exchange interaction and reduced field,
respectively. Taking the exchange interaction as the energy unit, the
dimensionless temperature will be denoted with $t=T/J_{\rm{c}}$, so
that all thermodynamic quantities depend upon the pair $(h,t)$.
Periodic boundary conditions are assumed
($\bm{s}_{N+1}\,{=}\,\bm{s}_1$) and the thermodynamic limit
($N\,{\to}\,\infty$) will be considered.

It is important to distinguish between the CIF and what is often
called the \emph{classical Ising model}, which is obtained from
Eq.~(\ref{e.HQIF}) with $H\,{=}\,0$, by regarding the quantum
operators $\hat{S}^x_i$ as classical discrete variables taking the
values $\pm{1/2}$. Results for the CIF are lacking, apart from the
zero-field case~\cite{Horiguchi1990,Minami1996,Minami1998}, probably
because, at variance with the QIF, the CIF does not allow for an
exact solution.

Writing the classical spin variables in terms of polar angles as
\begin{equation}
 \bm{s}_i\equiv\big(
 \sin\theta_i,\cos\theta_i\sin\varphi_i,\cos\theta_i\cos\varphi_i\big)~,
\end{equation}
the Hamiltonian \eqref{e.H} is expressed as
\begin{equation}
 \frac{\cal H}{J_{\rm{c}}}=-\sum_i\big(
 \sin\theta_i\sin\theta_{i+1}+2h\cos\theta_i\cos\varphi_i\big)~;
\label{e.Hthetaphi}
\end{equation}
its minimum corresponds to a translation-invariant configuration
$\big\{\bm{s}_i\,{=}\,\big(\sin\theta_{\rm{m}},0,\cos\theta_{\rm{m}}\big)\big\}$,
with
\begin{equation}
 \theta_{\rm m}(h) = \left\{
 \begin{array}{ll}
 \pm\cos^{-1}h & ~~~{\rm for}~ h\leq 1
 \\
 0 & ~~~{\rm for}~ h\geq 1
 \end{array}\right.~~~.
\end{equation}
The minimum energy per spin is
\begin{equation}
 u(h,0) = \left\{
 \begin{array}{ll}
 -1-h^2 & ~~~{\rm for}~ h\leq 1
 \\
 - 2h & ~~~{\rm for}~ h\geq 1
 \end{array}\right.
\end{equation}
and shows a singularity at $h\,{=}\,1$. The magnetization $m_z$ is
proportional to the field for $h\,{\leq}\,1$ and saturates for
$h\,{\geq}\,1$,
\begin{equation}
 m_z(h,0) = \left\{
 \begin{array}{ll}
 h & ~~~{\rm for}~ h\leq 1
 \\
 1 & ~~~{\rm for}~ h\geq 1
 \end{array}\right.~~~,
\end{equation}
so that the corresponding susceptibility is discontinuous at
$h\,{=}\,1$
\begin{equation}
 \chi_z(h,0)=\partial_h{m_z}= \left\{
 \begin{array}{ll}
 1 & ~~~{\rm for}~ h<1
 \\
 0 & ~~~{\rm for}~ h>1
 \end{array}\right.~~~.
\label{e.chi-t0}
\end{equation}
Finally, the behavior of the magnetization along the exchange,
\begin{equation}
 m_x(h,0) = \left\{
 \begin{array}{ll}
 \pm\sqrt{1-h^2} & ~~~{\rm for}~ h\leq 1
 \\
 0 & ~~~{\rm for}~ h\geq 1
 \end{array}\right.~~~,
\label{e.mx-t0}
\end{equation}
reflects the fact that for $h\,{<}\,1$ the minimum is twofold.

It may sound odd but, as seen in Fig.~1, the CIF displays a
zero-temperature behavior which is analogous to that observed in the
QIF, even if no fluctuations are present at $t\,{=}\,0$.
Eq.~\eqref{e.mx-t0} shows indeed that a \emph{critical} field,
$h_{\rm{c}}\,{=}\,1$, separates a symmetry-broken minimum-energy
configuration with $m_x\,{\neq}\,0$ from one with $m_x\,{=}\,0$.

An even closer analogy is found if one considers that from the
low-temperature expression derived in Appendix~\ref{a.SSWT} one can
obtain the exact zero-$t$ limit of the correlation length,
$\xi_x(h,0)=1/\sqrt{h{-}1}$. The counterparts of both
Eqs.~\eqref{e.Mxquantum} and \eqref{e.Xixquantum} are then available,
and read
\begin{equation}
  m_x\sim (h_{\rm{c}}-h)^{\beta}~~~{\rm for}~h\to h_{\rm{c}}^-~,
 \label{e.Mxclassical}
\end{equation}
and
\begin{equation}
  \xi_x\sim (h-h_{\rm{c}})^{-\nu}~~~{\rm for}~h\to h_{\rm{c}}^+~,
 \label{e.Xixclassical}
\end{equation}
with Gaussian critical exponents $\beta=\nu=1/2$, to be compared with
those for the QIF, $\beta=1/8$ and $\nu=1$.

\section{The Ising model in a transverse field : $T>0$}
 \label{s.ItF.T>0}
  \subsection{$T>0$ : The quantum model}
   \label{ss.QIF.T>0}

\begin{figure}
\includegraphics[width=86mm,angle=0,clip=true]{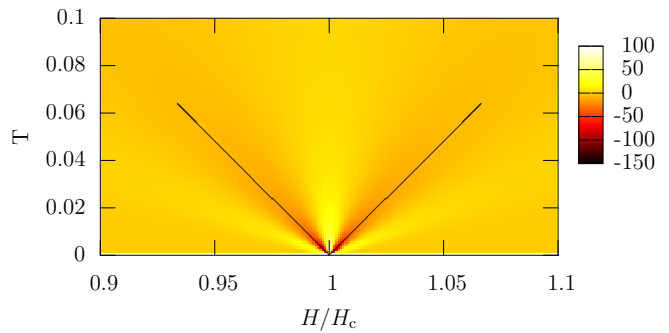}
\caption{Second derivative of the specific heath of the QIC,
$\partial^2_Hc(H,T)$.}
 \label{f.q-sh}
\end{figure}

The field- and temperature dependence of the specific heat $c(H,T)$
and of the susceptibility $\chi_z(H,T)$ can be easily obtained from
the analytic results of Ref.~\onlinecite{Pfeuty1970}. The most
prominent feature is the occurrence of maxima of both quantities in
the $H{-}T$ plane. Indeed, the quantum specific heat just shows the
behavior of a free Fermi gas with dispersion
$\omega_k\sim\Delta\,{+}\,k^2$ in the neighborhood of a vanishing gap
$\Delta\sim|H\,{-}\,H_{\rm{c}}|$, which appears, for fixed low-$T$,
as two symmetric peaks at linearly displaced positions
$|H\,{-}\,H_{\rm{c}}|\propto{T}$. This feature is made evident in
Fig.~\ref{f.q-sh}, by the density plot of the second derivative of
$c(H,T)$ with respect to $H$. The positions of the maxima draw two
symmetric lines in the $H-T$ plane which coincide with those
obtained~\cite{AmicoP2006} from the analysis of entanglement
properties.

  \subsection{$T>0$ : The classical model}
   \label{ss.CIF.T>0}

The thermodynamic behavior of the CIF for $h\,{<}\,h_{\rm{c}}$
is essentially determined by the energy landscape of the model. In a
mean-field approach, i.e., setting
$\{\theta_i\,{=}\,\theta,\varphi_i\,{=}\,0\}$, this is described by
the double-well energy profile
$e(\theta)=-\sin^2\theta\,{-}\,2h\cos\theta$, with minima in
$\theta_{\rm{m}}\,{=}\,\pm\cos^{-1}h$ and the barrier top at
$\theta\,{=}\,0$, the barrier energy being
$\delta{e}\,{=}\,(1\,{-}\,h)^2=\,(h_{\rm{c}}\,{-}\,h)^2$. The two
wells correspond to the Ising configurations.

Domain-wall excitations connecting the two Ising configurations can
appear on the chain, the energy of a domain wall being
$e_{\rm{w}}=2(1-h^2)$. A simple statistical argument gives a finite
number of domain walls at any finite temperature,
$n_{\rm{w}}\,{\sim}\,N/(1+e^{e_{\rm{w}}/t})$, so that when
temperature is switched on the ordered state is destroyed (i.e.,
$m_x\,{=}\,0$ for $t\,{>}\,0$) by these excitations, which rule the
low-temperature thermodynamics in what we will hereafter call the
\emph{Ising} regime.

When the temperature reaches the order of the barrier energy,
$t\,{\sim}\,(h_{\rm{c}}\,{-}\,h)^2$, thermally activated transitions
between the wells can occur and the Ising regime breaks down. The
dependence of the barrier height on the field is responsible for the
fact that as $h\to h_{\rm{c}}^-$ this regime gets confined into the
narrow interval $0<t<(h_{\rm{c}}-h)^2$. The above mechanism, that we
call \emph{thermal hopping}, is at the hearth of the phenomenology of
the model below $h_{\rm{c}}$, and it already suggests the occurrence
of a crossover from an Ising-like behavior towards a \emph{critical}
one, ruled by an effectively flat energy-landscape.

On this basis, let us discuss the finite-temperature data
obtained~\cite{Taiti2006} by the Transfer-matrix method, briefly
described in Appendix~\ref{a.TM}.

\begin{figure}
\includegraphics[height=84mm,angle=90]{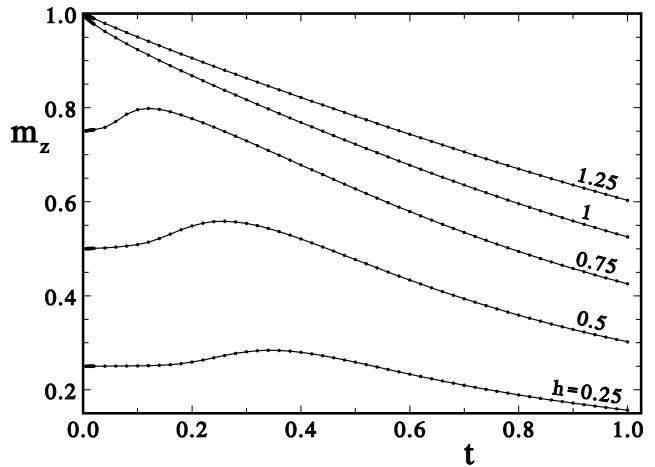}
\caption{Magnetization along the field direction, $m_z(h,t)$ \emph{vs}
temperature, for selected field values. The `critical'
value is $h_{\rm{c}}\,{=}\,1$.}
\label{f.mz-t}
\end{figure}

The temperature behavior of the magnetization along the field,
$m_z(h,t)$, is shown in Fig.~\ref{f.mz-t}, and clearly is related to
that of the mean-square fluctuations of the Ising order parameter,
$\langle(s_i^x)^2\rangle$; indeed, the initial increase with
temperature corresponds to the reduction of $\langle(s_i^x)^2\rangle$
due to the fact that the probability distribution of $s_i^x$,
initially frozen in the bottom of one of the two wells
($\pm\sqrt{1-h^2}$), extends more likely towards the barrier at
$s_i^x\,{=}\,0$. The further decay of $m_z$ is due to the isotropic
spin fluctuations occurring after thermal hopping has taken place. On
the other hand, for $h\,{\ge}\,h_{\rm{c}}$, $m_z$ simply decreases
from its $t\,{=}\,0$ saturation value.

\begin{figure}
\includegraphics[height=84mm,angle=90]{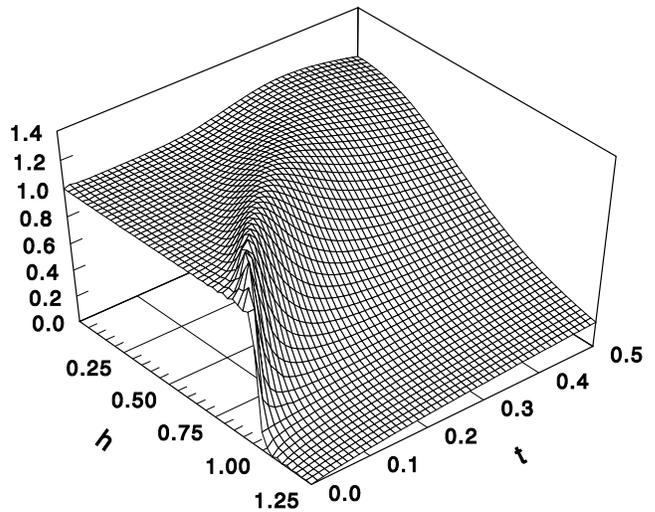}
\caption{Magnetic susceptibility $\chi_z(h,t)$.}
 \label{f.chi-ht}
\end{figure}

The susceptibility $\chi_z(h,t)$ is shown in Fig.~\ref{f.chi-ht}: the
zero-field result agrees with that derived in
Ref.~\onlinecite{Minami1996} and shows a broad maximum in
temperature, at $t\,{\simeq}\,0.37$. Upon rising the field, such
maximum is squeezed towards lower temperatures, meanwhile getting
sharper. At the critical field, the maximum disappears and the
susceptibility is a monotonic function of temperature for whatever
$h\,{\geq}\,h_{\rm{c}}$. The zero-$t$ limiting value of $\chi_z(h,t)$
is given by Eq.~\eqref{e.chi-t0}. The overall behavior of
$\chi_z(h,t)$ in the $h$--$t$ plane is evidently characterized by the
occurrence of the above described maxima for $h\,{<}\,h_{\rm{c}}$.

\begin{figure}
\includegraphics[height=84mm,angle=90]{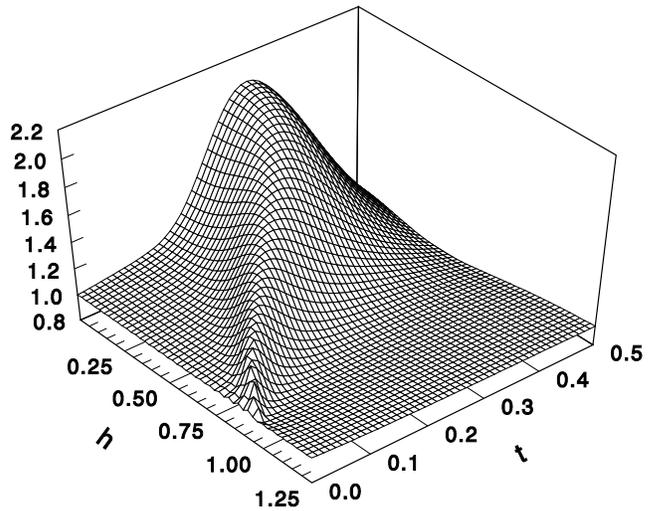}
\caption{Specific heat $c(h,t)$.}
 \label{f.c-ht}
\end{figure}

\begin{figure}
\includegraphics[height=84mm,angle=90]{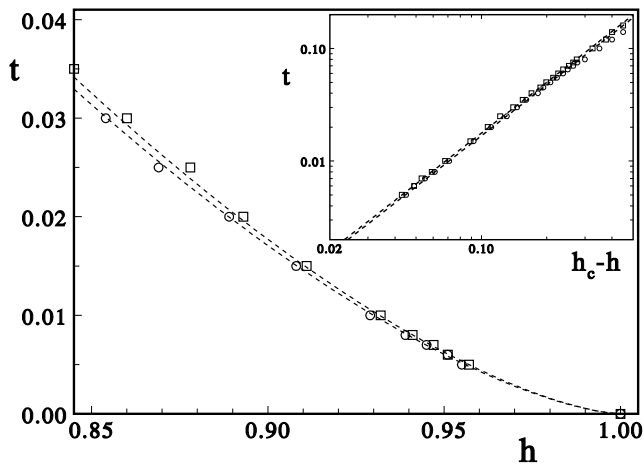}
\caption{
Region $h\lesssim{h}_{\rm{c}}$ of the phase diagram of the CIF.
Circles and squares indicate the position of the maxima of $c(h,t)$ and
$\chi(h,t)$, respectively. The dashed curves are obtained fitting the data
with $t\propto(1{-}h)^{3/2}$. Inset: log-log plot of the same data and
curves.}
 \label{f.hct}
\end{figure}

The specific heat $c(h,t)$, shown in Fig.~\ref{f.c-ht}, is also
characterized by the occurrence of maxima for $h<h_{\rm{c}}$, which
disappear above the critical field; noticeably they fall into almost
the same positions as those observed in $\chi_x(h,t)$, as it appears
in Fig.~\ref{f.hct}.

The maxima observed both in the susceptibility and in the specific
heat correspond to the onset of thermal hopping, and their positions
indicate the region where thermal fluctuations overcome the Ising
domain-wall energy, i.e., the crossover region from the \emph{Ising}
to the \emph{critical} regime. In order to better characterize the
corresponding crossover line we have fitted the maxima positions for
low $t$ in the $h$--$t$ plane with the function
$t\,{\propto}\,(h_{\rm{c}}{-}h)^\kappa$, finding the exponent quite
close to the value $\kappa\,{=}\,3/2$ derived by analytical arguments
in the next section.

\section{The phase diagram}
 \label{s.PhaseDiagram}
  \subsection{Quantum model}
   \label{ss.QuantumPhaseDiagram}

\begin{figure}
\includegraphics[height=84mm,angle=90]{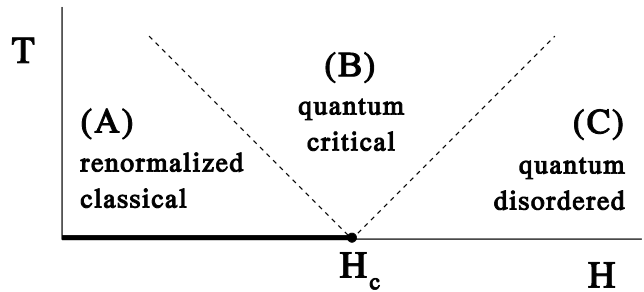}
\caption{
Phase diagram of the quantum Ising model in a transverse field.}
 \label{f.phd}
\end{figure}

The finite-temperature phase diagram for the QIF, shown in
Fig.~\ref{f.phd}, essentially features the occurrence of three
regions, characterized by qualitatively different behavior of
physical observables with respect to $h$ and~$t$. These regions have
been identified~\cite{ChakravartyHN1989} with the so called
\emph{renormalized-classical} (\textbf{A}),
\emph{quantum-critical} (\textbf{B}), and \emph{quantum-disordered}
(\textbf{C}) regimes, which have been singled out according to the
qualitatively different behavior of the correlation length $\xi_x$,
which is
\begin{equation}
 \xi_x \sim  \left\{
\begin{array}{lll}
 e^{u(H_{\rm{c}}{-}H)/T}~~~
      &\mbox{for}~~ H_{\rm{c}}-H \gg T & (\textbf{A})
\\[1mm]
 T^{-1/y} &\mbox{for}~ |H-H_{\rm{c}}|\ll T~~ & (\textbf{B})
\\[1mm]
 (H{-}H_{\rm{c}})^{-\nu} &\mbox{for}~~ H-H_{\rm{c}} \gg T & (\textbf{C})
\end{array}
\right.
\label{e.xiq}
\end{equation}
where $u$ is a constant and the critical exponents are $y\,{=}\,1$
and $\nu\,{=}\,1$. The power law divergence in
Eqs.~\eqref{e.xiq}-(\textbf{B}) and -(\textbf{C}) follow from the
temperature-dependent scaling law for $\xi_x$ lying at the hearth of
the Renormalization Group (RG) approach to critical
phenomena~\cite{Continentino2001,Sachdev1999}. We remind that the
exponent $y$, ruling the scaling of energy, is bound to equal the
dynamical critical exponent $z$ by the uncertainty principle. In the
same framework, the above regimes are shown to be separated by the
crossover lines $T\propto |H\,{-}\,H_{\rm{c}}|$.

This phase-diagram has been extensively discussed (see, e.g.,
Refs.~\onlinecite{Sachdev1999} and~\onlinecite{Continentino2001})
over the last decade: in particular, the fact that $\xi_x$ is
independent of the field and inversely proportional to $T$ in region
\textbf{B}, and independent of temperature and inversely proportional
to $H-H_{\rm{c}}$ in region
\textbf{C}, has always been considered as a signature of the genuinely
quantum character of the corresponding regimes, which have
consequently been labeled as ``quantum'' (critical and disordered,
respectively).

In fact, as we argue in the remaining part of this article, this is
not truly the case.

  \subsection{Classical model}
   \label{ss.ClassicalPhaseDiagram}

\begin{figure}
\includegraphics[height=84mm,angle=90]{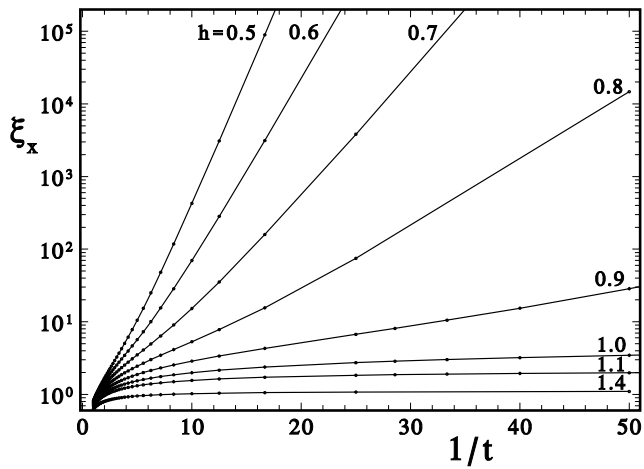}
\caption{Correlation length $\xi_x(h,t)$ \emph{vs} $1/t$, for
selected field values.}
 \label{f.xix-t}
\end{figure}

\begin{figure}
\includegraphics[height=84mm,angle=90]{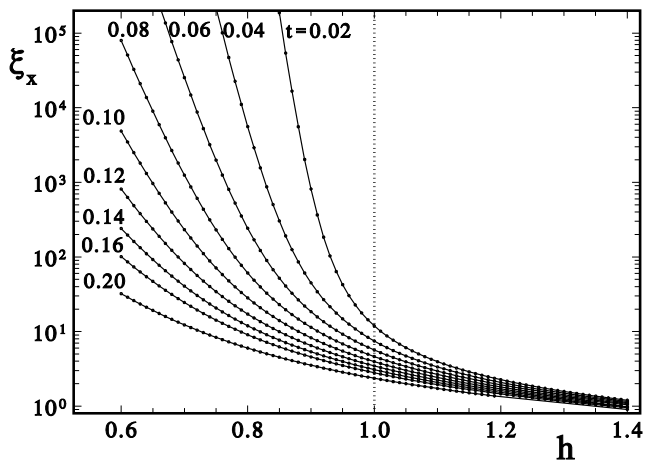}
\caption{Correlation length $\xi_x(h,t)$ \emph{vs} field, for
different temperatures.The vertical dotted line marks the critical
field $h_{\rm{c}}\,{=}\,1$.}
 \label{f.xix-h}
\end{figure}

As seen in Sec.~\ref{ss.CIF.T>0}, hints of the possible occurrence of
at least two different regimes, separated by a crossover region, do
already come from the behavior of the susceptibility and of the
specific heat for $h<h_{\rm{c}}$. However, in order to closely mimic
the procedure followed in drawing the quantum phase diagram, we
analyze the field and temperature dependence of the correlation
length $\xi_x(h,t)$ of the classical model, heading towards
expressions analogous to those of Eqs.~\eqref{e.xiq}. The overall
behavior of $\xi_x(h,t)$, as from our numerical data, is plotted in
Figs.~\ref{f.xix-t} and~\ref{f.xix-h}.


\subsubsection{Correlation length for $h<h_{\rm{c}}$}

\begin{figure}
\includegraphics[height=84mm,angle=90]{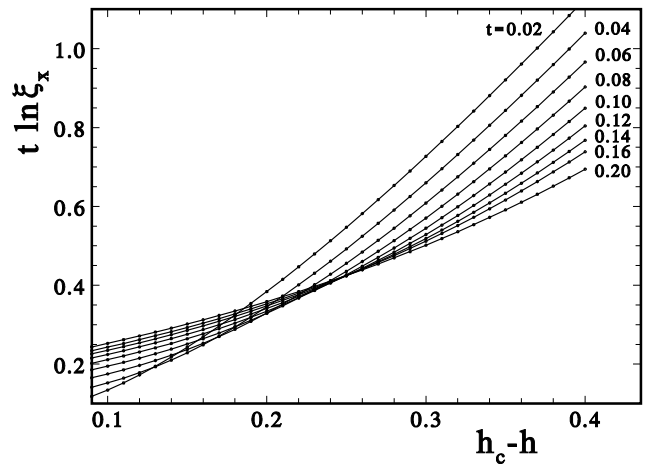}
\caption{Classical phase-diagram region \textbf{a}. The function
$t\ln\xi_x(h,t)]$ vs $h_{\rm{c}}-h$ for different fixed temperatures.
One can see that the curves become straight lines the better the
condition $h_{\rm{c}}{-}h\,{\gg}\,t$ is satisfied.}
 \label{f.tlnxi-h}
\end{figure}

If one considers the $t$-dependence for $h<h_{\rm{c}}$, it clearly
appears from Fig.~\ref{f.xix-t} that $\ln\xi_x\propto 1/t$, with a
slope that increases with the difference $h_{\rm{c}}\,{-}\,h$. As for
the field dependence, the logarithm of $\xi_x(h,t)$ is seen in
Fig.~\ref{f.xix-h} to be linear in $h$, with negative slope: this is
the more evident the smaller the temperature and the angular
coefficient decreases with increasing $t$, consistently with
Eq.~\eqref{e.xiq}-(\textbf{A}). In fact, from Fig.~\ref{f.tlnxi-h}
for $h_{\rm{c}}\,{-}\,h\gg{t}$ we see that $t\,\ln\xi_x(h,t)\approx
f(t)+a(h_{\rm{c}}\,{-}\,h)$, with $f(t)$ weakly dependent on $t$,
meaning
\begin{equation}
 \xi_x(h,t)\sim~ e^{u(h_{\rm{c}}\,{-}\,h)/t}
 ~~~{\rm for}~~h_{\rm{c}}-h\gg t~~~(\textbf{a}).
\label{e.xicA}
\end{equation}

\subsubsection{Correlation length for $h\ge h_{\rm{c}}$}

From Figs.~\ref{f.xix-t} and~\ref{f.xix-h} we see that for field
close and above the critical value, the temperature- and field
dependence of $\xi_x(h,t)$ becomes much less pronounced with respect
to those displayed below the critical field, suggesting power-law
behaviors of the same type observed in the quantum regimes \textbf{B}
and \textbf{C}.

\begin{figure}
\includegraphics[height=84mm,angle=90]{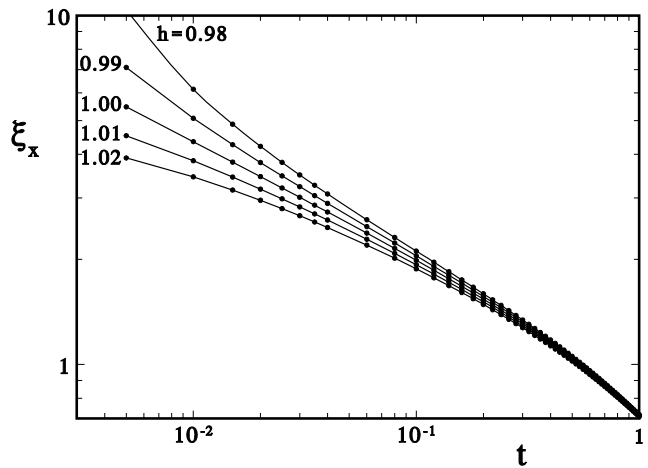}
\caption{Classical phase-diagram region \textbf{b}. Longitudinal
correlation length $\xi_x(h,t)$ vs $t$ for fields close to the
critical value $h_{\rm{c}}\,{=}\,1$. The log-log plot emphasizes that
the slope of the curve for $h\,{=}\,h_{\rm{c}}$ is -1/3.}
 \label{f.xix-t-hc}
\end{figure}

\begin{figure}
\includegraphics[height=84mm,angle=90]{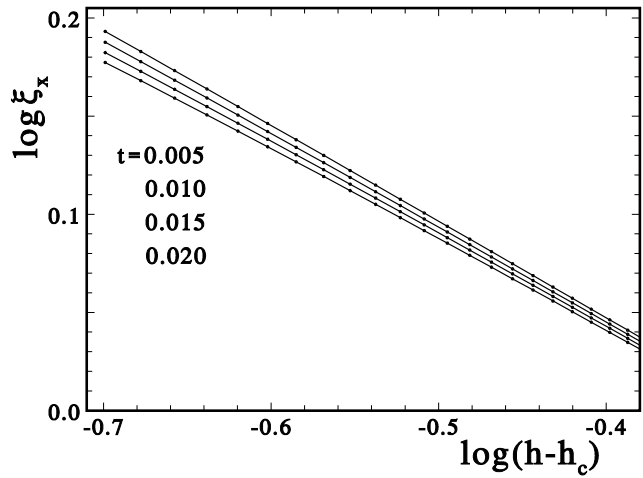}
\caption{Classical phase-diagram region \textbf{c}. Log-log plot of
$\xi_x(h,t)$ vs $h{-}h_{\rm{c}}$, for different low temperatures. The
slope of the low-$t$ curves is seen to be -1/2.}
 \label{f.xix-h0}
\end{figure}

In particular, using our numerical data, we can ascertain that, for
$|h-h_{\rm{c}}|\ll t$ the correlation length behaves as $\xi_x\propto
t^{-1/3}$, as evidenced in Fig.~\ref{f.xix-t-hc}; on the other hand,
for $|h-h_{\rm{c}}|\gg t$, the log-log plot reported in
Fig.~\ref{f.xix-h0} emphasizes a power-law field-dependence,
$\xi_x\,{\propto}\,(h-h_{\rm{c}})^{-1/2}$. These behaviors are fully
analogous to the quantum ones described by
Eq.~\eqref{e.xiq}-(\textbf{B}) and~-(\textbf{C}), the only difference
being in the exponents.

In order to strengthen this result, we develop the analytical
treatment reported in Appendix~\ref{a.SSWT}, which makes use of
self-consistent spin-wave theory (SSWT) for $h\,{\geq}\,h_{\rm{c}}$.
The asymptotic behavior of the correlation length derived there,
valid at low temperature and close to criticality, is given by
Eq.~\eqref{e.xiDelta} and yields
\begin{equation}
 \xi_x(h,t)\sim t^{-1/3}
 ~~~~~~~~~~{\rm for}~~|h-h_{\rm{c}}|\ll t ~~~(\textbf{b}),
\label{e.xicB}
\end{equation}
and
\begin{equation}
 \xi_x(h,t)\sim (h-h_{\rm{c}})^{-1/2}
 ~~~{\rm for}~~|h-h_{\rm{c}}|\gg t~~~(\textbf{c}),
\label{e.xicC}
\end{equation}
in full agreement with the analysis of our numerical data.


\subsubsection{Crossovers}

Let us now consider the problem of identifying the crossover regions
between phases where Eqs.~\eqref{e.xicA},~\eqref{e.xicB},
and~\eqref{e.xicC} hold, i.e., between the three classical regimes,
that we have labeled \textbf{a}, \textbf{b}, and \textbf{c},
respectively.

As for the \textbf{ab} crossover, a reasonable localization can be
obtained by relating it with thermal hopping. According to the
mean-field analysis presented in Sec.~\ref{ss.CIF.T>0}, the latter
occurs when the temperature overcomes the energy-barrier height,
i.e., for $h_{\rm{c}}-h\sim{t}^{1/2}$. However, the mean-field
approach neglects correlated fluctuations, while to get a correct
estimate it is necessary to keep nearest-neighbor fluctuations at
least within a quadratic approximation of the
Hamiltonian~\eqref{e.Hthetaphi} around one of the minima. Setting
$\theta_i\,{=}\,\theta_{\rm{m}}\,{+}\,\varepsilon_i$ and expanding,
one finds for the quadratic part
\begin{equation}
 \frac{\cal H}{J_{\rm{c}}}\simeq N\,e(\theta_{\rm{m}})
 +h^2\sum_k\varphi_k^2 +\sum_k(1-h^2\cos k)\,\varepsilon_k^2~,
\end{equation}
from which the mean square fluctuation of $\theta_i$ around the
minimum results
\begin{equation}
 \langle\varepsilon_i^2\rangle
 \simeq\frac t{2N}\sum_k \frac1{1-h^2\cos k}
 = \frac t{2\sqrt{1-h^4}} ~.
\end{equation}
As soon as these fluctuations reach the size of the width of the
barrier, $|\theta_{\rm{m}}|=\sin^{-1}\sqrt{1-h^2}$, the Ising
excitations disappear and the crossover between the exponential- and
the power-law correlation length is expected. This condition is
fulfilled when
$\langle\varepsilon_i^2\rangle\simeq|\theta_{\rm{m}}|^2$, i.e., for
$h$ close to the critical field $h_{\rm{c}}\,{=}\,1$,
\begin{equation}
 {t}\sim(h_{\rm{c}}-h)^{3/2} \hspace{10mm}(\textbf{ab})~.
\label{e.clcrAB}
\end{equation}
This prediction gives the correct exponent and is also fully
consistent with that obtained in Sec.~\ref{ss.CIF.T>0} by fitting the
positions of the maxima observed, in the $h-t$ plane, for both the
susceptibility and the specific heat.

It is worth mentioning that thermal hopping is the ultimate cause of
the finite-temperature phase transition to the ordered state
occurring for $h\,{<}\,h_{\rm{c}}$ in more than one dimension: the
crossover turns indeed into the sharp critical line $t_{\rm{c}}(h)$,
with the critical temperature vanishing at the quantum critical
point, $t_{\rm{c}}(h_{\rm{c}})=0$. In one dimension the onset of
ordering is forbidden at finite $t$ and only the broad crossover
between regimes \textbf{a} and \textbf{b} survives.

Coming to the \textbf{bc} crossover, from Eqs.~\eqref{e.xicB}
and~\eqref{e.xicC} one deduces it to occur for
\begin{equation}
 {t}\sim (h-h_{\rm{c}})^{3/2}\hspace{10mm}(\textbf{bc})~.
\label{e.clcrBC}
\end{equation}

\subsubsection{Phase diagram}
Eqs.~\eqref{e.xicA},~\eqref{e.xicB}, and~\eqref{e.xicC}, can be
summarized as:
\begin{equation}
 \xi_x\sim \left\{
\begin{array}{lll}
 e^{u(h_{\rm{c}}{-}h)/t} & \mbox{for}~~h_{\rm{c}}-h \gg t & (\textbf{a})
\\[1mm]
 t^{-1/y} &\mbox{for}~ |h-h_{\rm{c}}|\ll t~~ & (\textbf{b})
\\[1mm]
 (h-h_{\rm{c}})^{-\nu}~~ &\mbox{for}~~ h-h_{\rm{c}} \gg t & (\textbf{c})
\end{array}
\right.
\label{e.xicl}
\end{equation}
with $y\,{=}\,3$ and $\nu\,{=}\,1/2$. Via the different behavior of
the correlation length, three different regimes can thus be singled
out also in the classical phase diagram and, according to our
discussion, it makes sense to call them \emph{Ising} (\textbf{a}),
\emph{critical} (\textbf{b}), and \emph{disordered} (\textbf{c}).
Different regimes are again separated by crossover lines, which in
the classical model are described by the relation
$t\,{\sim}\,|h\,{-}\,h_{\rm{c}}|^{3/2}$.

\subsection{Classical vs quantum phase diagram}
\label{ss.cvsq}

Eqs.~\eqref{e.xicl} clearly show that the CIF is characterized by a
phase diagram on the $h-t$ plane which is fully analogous to the
celebrated quantum one described by Eqs.~\eqref{e.xiq}. In
particular, the observed power-law divergence of $\xi_x$ in regimes
({\bf b}) and ({\bf c}) suggests the scaling hypothesis to hold also
in the classical case: This is confirmed in Appendix \ref{a.SSWT}
where we obtain an explicit expression for $\xi_x(h,t)$, which turns
out to be a homogeneous function.

The nexus between the quantum and the classical case can be drawn as
follows: Let us introduce the unifying parameters $\overline g$ and
$\overline t$, defined as $\overline g=H-H_c$ ,
$\overline t=T/J$ in the quantum model, and $\overline g=h\,{-}\,h_c$,
$\overline t=t$ in the classical model, and write the equation
\begin{equation}
 \xi_x(\overline g, \overline t)=
b\,\xi_x\big(b^{1/\nu}\overline g, b^y\,\overline t\big)
\label{e.xiscal}
\end{equation}
which, within RG, rules the scaling of observable quantities and
model parameters in the proximity of critical points after a
length-scale transformation by a factor $b$. From
Eq.~\eqref{e.xiscal} follow both the quantum
Eqs.~\eqref{e.xiq}-(\textbf{B}) and -(\textbf{C}) and the classical
ones,~\eqref{e.xicl}-(\textbf{b}) and -(\textbf{c}), as well as the
crossover lines $\overline{t}\propto{\overline g}^{\nu{y}}$: The
critical exponents entering the above expressions, despite getting
different values in the quantum ($y=1,\nu=1$) and in the classical
($y=3,\nu=1/2$) case, consistently fulfil the hyperscaling relation
\begin{equation}
2-\alpha=\nu(d+y)~,
\label{e.hyper}
\end{equation}
where $2\,{-}\,\alpha$ is the exponent for free-energy density $f$,
defined~\cite{Continentino2001} by
$f(\overline{g},0)\,{\sim}\,|\overline{g}|^{2-\alpha}$ ($\alpha=0$ in
both cases). We remind that, in the quantum case, the scaling
exponent of energy $y$ is unavoidably related to the dynamical
critical exponent $z$ by the uncertainty relation, $y\,{=}\,z$. As
for the classical case, we notice that $\nu$ takes the typical
Gaussian value, due to fluctuations freezing as $t\to{0}$.

\section{Conclusions}
\label{s.concl}

In this paper we have compared the finite-temperature phase diagram
of a quantum model displaying a QPT, namely the $S=1/2$ Ising chain
in a transverse field, with that of its classical limit. To this end,
we have obtained numerical and analytical results for the classical
model. In particular, we have studied the magnetizations, the
specific heat, the magnetic susceptibility, and the correlation
length along the exchange direction: all quantities have been
analyzed below, at, and above the saturation field, with the
temperature raised from zero up to values of the order of the
exchange interaction.

The classical phase diagram emerging from our work is fully analogous
to that of the quantum model. Three regimes are identified: the
\emph{Ising} regime \textbf{a} ($h_{\rm{c}}{-}h\,{\gg}\,t$),
corresponding to the quantum \emph{renormalized classical} regime,
where $\xi_x$ behaves exponentially with $(h_{\rm c}{-}h)/t$; the
\emph{critical} regime \textbf{b} ($|h{-}h_{\rm{c}}|\,{\ll}\,t$),
corresponding to the \emph{quantum critical} regime, where $\xi_x$
behaves algebraically with $t$; the \emph{disordered} regime
\textbf{c} ($h{-}h_{\rm{c}}\,{\gg}\,{t}$), corresponding to the
\emph{quantum disordered} regime, where $\xi_x$ behaves
algebraically with $h{-}h_{\rm{c}}$. Two crossover lines,
$t\,{\sim}\,|h{-}h_{\rm{c}}|^{3/2}$, separate the regions of the
phase diagram where the above regimes occur.

The essential message of this work is that in order to discriminate
quantum critical effects it is not sufficient to observe, say, an
algebraic behavior of the correlation length with respect to
temperature, but a precise determination of the exponent is rather
due. Our analysis does also suggest that the role of genuinely
quantum fluctuations at finite temperature is not as relevant as
commonly believed, given the fact that most of the features regarded
as typical of the quantum model are disclosed also in its classical
limit, even, and most noticeably, at very low temperature.

Given the very weak model-dependence of the overall discussion, we believe
that the above conclusions hold in general, and not only for the Ising
chain in a transverse field.

\medskip

We gratefully acknowledge fruitful discussions with A. Fubini. This
work was supported by MIUR under the 2005-2007 PRIN--COFIN National
Research Projects Program, n. 2005029421\_004.

\appendix

\section{Transfer-matrix for the CIF}
\label{a.TM}

We outline here the numerical Transfer-matrix
technique,~\cite{BlumeHL1975} by which we have investigated different
static thermodynamic quantities of the CIF, focusing the attention on
their low-temperature behavior in the neighborhood of the point
$h=h_{\rm{c}}$.

Using polar coordinates one can map the classical spins appearing in
the Hamiltonian~\eqref{e.H} as
\begin{equation}
 \textstyle
 \bm{s}_i=\big(
 x_i,\sqrt{1-x_i^2}\sin\varphi_i,\sqrt{1-x_i^2}\cos\varphi_i\big)~,
\end{equation}
with $x_i\in[-1,1]$ and $\varphi_i\in[0,2\pi]$. The partition
function can then be expressed as the trace of the $N$-th power of an
integral kernel $K(x,y)$,
\begin{eqnarray}
 {\cal Z} &=&
 \prod_{i=1}^N \int_{-1}^1\!\! \frac{dx_i}2 \int_{-\pi}^\pi
 \frac{d\varphi_i}{2\pi}
 \,e^{(x_ix_{i+1}+2h\sqrt{1-x_i^2}\cos\varphi_i)/t}
\notag\\
 &=& \int_{-1}^1\!\! dx~K^N(x,x) = \sum_\ell \lambda_\ell^N~,
\end{eqnarray}
where the kernel is real and symmetric,
\begin{eqnarray}
 K(x,y) &=& \frac{e^{xy/t}}2
 \Big[I_0\Big({\textstyle \frac{2h}t}\sqrt{1{-}x^2}\,\Big)
      I_0\Big({\textstyle \frac{2h}t}\sqrt{1{-}{y}^2}\,\Big)\Big]^{\frac12}
\notag\\
 &=& \sum_\ell \lambda_\ell\,\psi_\ell(x)\psi_\ell(y)~,
\end{eqnarray}
$I_0(x)$ is the modified Bessel function,
$\{\lambda_\ell\}=\{\lambda_0,\lambda_1,...\}$ are the (positive)
eigenvalues of $K$ (say, in decreasing order), and $\{\psi_\ell(x)\}$
the corresponding (real) eigenfunctions. The diagonalization of $K$
was performed numerically after discretizing the integral with a
5-point Simpson's formula~\cite{AbramowitzS1964} on a mesh of up to
1040 intervals, also accounting for the definite parity of the
eigenfunctions.

In the thermodynamic limit the free energy per site is a function of
the largest eigenvalue only,
\begin{equation}
 f(h,t) = -t \lim_{N\to\infty}\frac1N\ln{\cal Z}_N
 =-t~\ln\lambda_0(h,t)~;
\end{equation}
the internal energy $u\,{=}\,t^2\partial_t\ln\lambda_0$ and the
specific heat $c\,{=}\,\partial_t{u}$, were obtained by numerical
differentiation (5-point Lagrange formula~\cite{AbramowitzS1964}).

The probability distribution for the variable $x=x_i$ is
\begin{equation}
 w(x) = \psi_0^2(x)~;
\end{equation}
one can reduce to averages with this probability both the expressions
for the magnetization along the field
$m_z\,{=}\,\langle{s^z_i}\rangle\,{=}\,(t/2)\partial_h\ln\lambda_0$
and for the corresponding susceptibility
$\chi_z\,{=}\,\partial_h{m_z}$. On the other hand, the joint
probability for two sites at a distance $r$, $x=x_i$ and $y=x_{i+r}$,
is
\begin{equation}
 w_r(x,y) = \sum_\ell
 \Big(\frac{\lambda_\ell}{\lambda_0}\Big)^r
 \psi_0(x)\psi_\ell(x)\psi_0(y)\psi_\ell(y).
\end{equation}
Using this result, one can write the correlation function of the spin
components in the direction of the exchange as
\begin{equation}
 \langle s^x_is^x_{i+r}\rangle = \sum_\ell
 \bigg[\int dx~ x\,\psi_0(x)\psi_\ell(x)\bigg]^2
 \Big(\frac{\lambda_\ell}{\lambda_0}\Big)^r~;
\end{equation}
as the eigenfunction $\psi_0(x)$ is even, for large $r$ the leading
term is that for $\ell\,{=}\,1$,
\begin{equation}
 \langle s^x_is^x_{i+r}\rangle ~ \mathop{\sim}_{r\to\infty}~
 \Big(\frac{\lambda_1}{\lambda_0}\Big)^r~,
\end{equation}
so that the corresponding correlation length is given by
\begin{equation}
 \xi_x (h,t) = \big[\ln(\lambda_0/\lambda_1)\big]^{-1}~.
\end{equation}

\section{Self-consistent spin-wave theory}
\label{a.SSWT}

For $h\geq h_{\rm{c}}$ the minimum configuration of the classical
Hamiltonian Eq.~\eqref{e.Hthetaphi} corresponds to the saturation
one, i.e., $\big\{\theta_i\,{=}\,0,\varphi_i\,{=}\,0\big\}$. In order
to estimate the low-temperature behavior of the correlation length,
we use self-consistent spin-wave theory (SSWT), i.e., we assume a
Gaussian distribution $\rho_0\,{=}\,e^{-\beta{\cal{H}}_0}$ in terms
of a trial quadratic Hamiltonian ${\cal{H}}_0$ whose coefficients are
self-consistently determined by requiring the identity of the
$\rho_0$ averages of ${\cal{H}}$ and ${\cal{H}}_0$, as well as of
their first and second derivatives.

In terms of the relevant Gaussian variances
$D\,{=}\,\langle\theta_i^2\rangle_0$,
$D'\,{=}\,\langle\theta_i\theta_{i+1}\rangle_0$,
$E\,{=}\,\langle\varphi_i^2\rangle_0$ (by symmetry
$\langle\varphi_i\vartheta_j\rangle_0\,{=}\,0$), the SSWT amounts to
set
\begin{eqnarray*}
 \sin\theta_i\sin\theta_{i+1} &=&
 \frac12\big[\cos(\theta_i{-}\theta_{i+1})
 -\cos(\theta_i{+}\theta_{i+1})\big]
\\
 &\simeq& e^{-D}\Big[(1{+}D)\sinh D'-D'\cosh D'
\\
 && - \sinh D'\,\frac{\theta_i^2+\theta_{i+1}^2}2
 +\cosh D'\,\theta_i\theta_{i+1})\Big],
\end{eqnarray*}
and
\begin{eqnarray*}
 \cos\theta_i\cos\varphi_i &=&
 \frac12\big[\cos(\theta_i{-}\varphi_i)
 +\cos(\theta_i{+}\varphi_i)\big]
\\
 &\simeq& e^{-F}\Big(1{+}F-\frac{\theta_i^2+\varphi_i^2}2\Big),
\end{eqnarray*}
where $F\equiv(D+E)/2$ and we used the SSWT identity
$\cos{x}\simeq{e}^{-\langle{x^2}\rangle/2}
\big[1+\frac12(\langle{x^2}\rangle-x^2)\big]$.

The SSWT Hamiltonian is diagonal in Fourier space:
\begin{equation}
 \frac{{\cal H}_0}{J_{\rm{c}}}=-Ne_0(t) + \sum_k\big[
 (A+B\,\mu_k)\,|\theta_k|^2 + C\, |\varphi_k|^2\big]~,
\label{e.Hscha}
\end{equation}
with $\mu_k=1-\cos{k}$~ and
\begin{eqnarray}
 A &=& h\,e^{-F}-e^{-D-D'}
\notag\\
 B &=& e^{-D}\cosh D'
\notag\\
 C &=& h\,e^{-F},
\label{e.abc}
\end{eqnarray}
while $e_0(t)$ collects the uniform contributions. It follows that
the self-consistent expressions for the variances are
\begin{eqnarray}
 D &=& \frac t{2N}\sum_k \frac1{A+B\,\mu_k} = \frac t{2\sqrt{A(A+2B)}}
\notag\\
 D' &=& \frac t{2N}\sum_k \frac{\cos k}{A+B\,\mu_k}
 = D-\frac t{2N}\sum_k \frac{\mu_k}{A+B\,\mu_k}
\notag\\
 E &=& \frac t{2\,C}~.
\label{e.DDE}
\end{eqnarray}

The stability condition for ${\cal{H}}_0$ is $A\geq{0}$, which
defines a threshold field,
\begin{equation}
 h_0(h,t)=e^{F-D-D'}<1~,
\end{equation}
above which the SSWT is meaningful. Note that the present approach
can describe the finite-temperature behavior of the system also for
$h\,{\lesssim}\,h_{\rm{c}}\,{=}\,1$, because the configuration
density can still be approximated by a Gaussian centered in
$\big\{\theta_i\,{=}\,0,\varphi_i\,{=}\,0\big\}$ as long as thermal
fluctuations are large enough to overcome the barrier between the two
symmetric minima $\theta_i\,{=}\,\pm\theta_{\rm{m}}$ of ${\cal{H}}$,
as explained in Sec.~\ref{ss.CIF.T>0}.

From the SSWT Hamiltonian~\eqref{e.Hscha} the Fourier transform of
the correlation function is immediately found,
\begin{equation}
 G_x(k)=\big\langle|\theta_k|^2\big\rangle_0
 =\frac t{2(A+B\,\mu_k)} ~,
\end{equation}
and can be used to evaluate the correlation length $\xi_x$,
\begin{equation}
 \xi_x^2=-\frac{G''_x(k)}{2\,G(k)}\,\bigg|_{k{=}0}
 = \frac{B}{2\,A}
 = \frac{e^{F-D}\cosh D'}{h-h_0(h,t)}~.
\label{e.xiSSWT}
\end{equation}

Therefore, when the field is close to the critical value
$h_{\rm{c}}\,{=}\,1$, the behavior of the correlation length is given
by $\xi_x\sim [h-h_0(h,t)]^{-1/2}$. In this region the variance $D$
is enhanced ($A\,{\sim}\,{0}$) and it can be easily seen seen that
$D'\simeq{D}$ and $F\simeq{D/2}$, so that
\begin{eqnarray}
 \Delta(h,t) &\equiv& 1-h_0(h,t) \simeq D+D'-F \simeq \frac32 D
\notag \\
 &\simeq& \frac{3\,t}{4}
 \big[(h-1+\Delta)(h+1+\Delta-D) \big]^{-1/2}
\notag \\
 &\simeq& c~t~(g+\Delta)^{-1/2}~,
\label{e.Delta}
\end{eqnarray}
with $g\equiv{h}\,{-}\,1$ and the constant $c=3/4\sqrt{2}$.

Rewriting this equation as
\begin{equation}
 \frac\Delta{t^{2/3}}\simeq
 c~\bigg(\frac{g}{t^{2/3}}+\frac\Delta{t^{2/3}}\bigg)^{-1/2}
\end{equation}
it appears that $\Delta/t^{2/3}\,{=}\,F(x)$, with $x\,{=}\,g/t^{2/3}$
and the asymptotic behaviors $F(0)\,{=}\,c^{2/3}$,
$F(x{\to}\infty)\,{\sim}\,c\,x^{-1/2}$. From Eqs.~\eqref{e.xiSSWT}
and~\eqref{e.Delta} the leading behavior of the correlation length is
given by
\begin{equation}
 \xi_x\sim (g+\Delta)^{-1/2} \simeq {t}^{-1/3}\,F(x)~,
\label{e.xiDelta}
\end{equation}
and it follows that near the critical point it is a homogeneous
function,
\begin{equation}
 \xi_x(b^2g,b^3t) \simeq b^{-1}\xi_x(g,t) ~,
\end{equation}
which coincides with Eq.~\eqref{e.xiscal} with the exponents
$\nu\,{=}\,1/2$ and $y\,{=}\,3$. For the crossover line
(\textbf{bc}), identified by imposing $x\,{\simeq}\,1$, one finds
$t\,{\sim}\,g^{3/2}$.

\end{document}